\documentclass[conference, a4paper]{IEEEtran}

\usepackage{times}
\usepackage{graphicx}
\usepackage{amsmath}   
\usepackage{amssymb}
\usepackage{epstopdf}
\graphicspath{ {./} }
\usepackage[hyphens]{url}
\usepackage{subfigure}
\usepackage{enumitem}
\usepackage{algorithm} 
\usepackage{algpseudocode}

\usepackage{subfigure}
\usepackage{multirow}

\usepackage{dsfont}
\usepackage{booktabs}
\usepackage{colortbl}
\usepackage{tabulary}
\usepackage{hyperref}

\renewcommand\IEEEkeywordsname{Keywords}

\begin{document}



\title{Continuous Authentication of Smartphones Based on Application Usage}


\author{\IEEEauthorblockN{Upal Mahbub\IEEEauthorrefmark{1}, Jukka Komulainen\IEEEauthorrefmark{2}, Denzil Ferreira\IEEEauthorrefmark{2} and Rama Chellappa\IEEEauthorrefmark{3}}
\IEEEauthorblockA{\IEEEauthorrefmark{1}\IEEEauthorrefmark{3}Department of Electrical and Computer Engineering and the Center for Automation Research, \\UMIACS, University of Maryland, College Park, MD 20742\\
}
\IEEEauthorblockA{\IEEEauthorrefmark{2}University of Oulu, Finland\\
Email: umahbub@umiacs.umd.edu, yty@iki.fi, denzil.ferreira@oulu.fi, rama@umiacs.umd.edu}
}

\maketitle

\begin{abstract}
An empirical investigation of active/continuous authentication for smartphones is presented in this paper by exploiting users' unique application usage data, i.e., distinct patterns of use, modeled by a Markovian process. Variations of Hidden Markov Models (HMMs) are evaluated for continuous user verification, and challenges due to the sparsity of session-wise data, an explosion of states, and handling unforeseen events in the test data are tackled. Unlike traditional approaches, the proposed formulation does not depend on the top N-apps, rather uses the complete app-usage information to achieve low latency. Through experimentation, empirical assessment of the impact of unforeseen events, i.e., unknown applications and unforeseen observations, on user verification is done via a modified edit-distance algorithm for simple sequence matching. It is found that for enhanced verification performance, unforeseen events should be incorporated in the models by adopting smoothing techniques with HMMs. For validation, extensive experiments on two distinct datasets are performed. The marginal smoothing technique is the most effective for user verification in terms of equal error rate (EER) and with a sampling rate of $1/30s^{-1}$ and $30$ minutes of historical data, and the method is capable of detecting an intrusion within $\sim2.5$ minutes of application use. 
%
%

%
%

\end{abstract}

\begin{IEEEkeywords}
Active authentication; application usage-based verification; unforeseen observation handling; hidden markov models; marginal smoothing; markov chains; sequence matching;
\end{IEEEkeywords}

\section{Introduction}
With the rapid increase of smartphone users worldwide, the mobile applications are growing both in number and popularity \cite{DiverseUsageBehaviorOfApps_Xu2011}. The number of apps in Google Play store is around 8 million, while in Apple App Store, Windows Store and Amazon Appstore there around 2.2 million, 669 thousand, and 600 thousand applications, respectively\footnote{\url{https://www.statista.com/statistics/276623/number-of-apps-available-in-leading-app-stores/}}. It has been estimated that a total of 197 billion mobile application were downloaded in 2017\footnote{\url{https://www.statista.com/statistics/271644/worldwide-free-and-paid-mobile-app-store-downloads/}}. A retrospective study in 2016 showed that on average a smartphone user uses over $30$ different mobile applications per month and $\sim 10$ different applications per day \cite{AppUsageStat_AppAnnie2017}. As for usage duration in 2016, in the USA, the smartphone users spend on a daily basis over two hours on mobile applications, \textit{i.e.}, over a month usage of applications in a year \cite{AppUsageStat_AppAnnie2017}. With growing concerns of smartphone security, monitoring the application usage coupled with the diverse pool of applications can help to make a difference in user authentication systems.  

\begin{figure}[t]
\centering
\includegraphics[width = 0.45\textwidth]{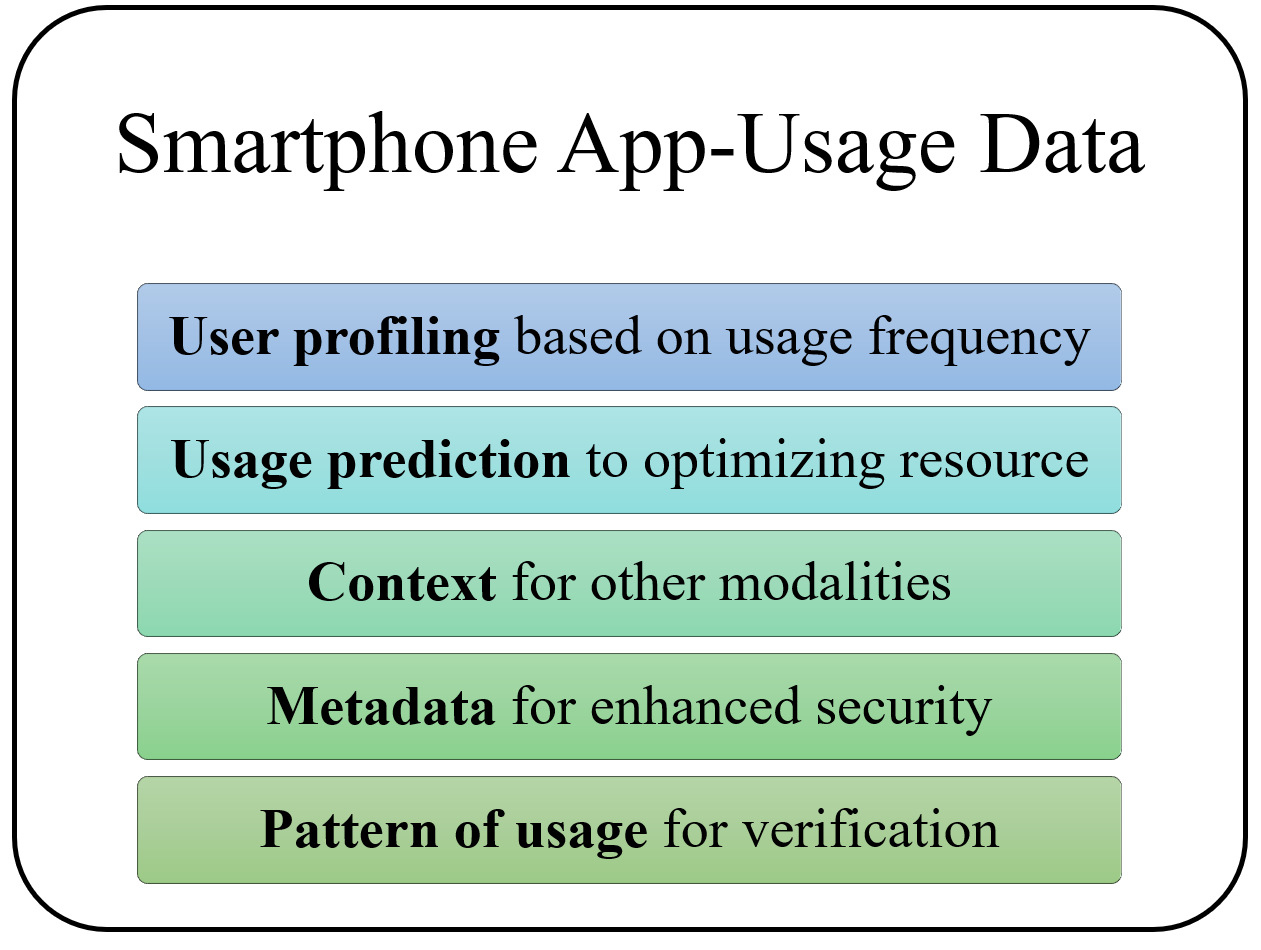}
\caption{Use cases for smartphone app-usage data.}
\label{AppDataUtilize}
\end{figure}

Smartphone application usage data can provide several interesting insights on the device users leading to different use cases of such data. There are several research works on user profiling and predicting behavioral patterns using application usage data \cite{PredictUsagePattern_Xu2013}\cite{DiverseUsageBehaviorOfApps_Xu2011}\cite{MobileMiner_Srinivasan2014}\cite{DiffUsersByApps_Welke2016}. Predicting application usage pattern can also help optimizing smartphone resources and help simulating realistic usage data for automated smartphone testing \cite{BatteryLifefromUsage_Houran2018}\cite{RevisitationAnalysis_Jones2015}\cite{Kostakos_UsageModeling}\cite{HumanBatteryInteraction_Ferreira2013}\cite{Testdroid_Kaasila2012}\cite{DiversityInUsage_Falaki2010}. The open foreground application can also work as a context for active authentication using other modalities \cite{ContextAwareTouch_Feng2014}\cite{AppCentricAuth_Khan2014}\cite{Mondal2015}\cite{Murmuria2015}. For example, when verifying with touch and accelerometer data, the application running in the background can provide useful context for robust authentication. Intuitively, the way a user handles and swipes in a phone for a banking application is very different from those for a gaming application. The foreground application context can be even more useful for active authentication if some more insightful information about the applications are available as metadata. For example, one key idea of active/continuous authentication is gradually blocking a probable intruder starting from the most sensitive applications, such as banking and social media accounts \cite{ProgressiveAuth_Riva2012} \cite{AA02_MahbubChellappa_BTAS2016}. If the sensitivity level or the type of application is known as metadata, it would be possible to attain enhanced security. Also, some applications, if permitted, can access the location data and store click information  for targeted advertisement and similar applications \cite{TargettedAd_Yang2016}.  A more active use case of application-usage data could be verifying the users solely from the pattern of usage. The different use cases of app-usage data are shown in Fig. \ref{AppDataUtilize}.
%
%
%
%
%
%
%
%

In this paper, the suitability of application-usage data as a modality for smartphone user verification is thoroughly investigated. The main contributions of this paper are:
\begin{itemize}
\item An innovative formulation that utilizes application usage data pattern as a biometric for user verification. The formulation tackles key challenges such as data sparsity and accounting for unforeseen test observations. Unlike traditional approaches of using top N-applications for authentication purposes \cite{AA_StylometryAppWeb_Friedman}, in the proposed formulation the full list of applications are considered for verification models in order to ensure low-latency which is essential for active authentication systems.
\item Insight into the application usage similarity among different users and statistics on unforeseen applications.
\item A thorough investigation of the impact of unknown applications and unforeseen observations on the verification task.  
\item A Modified Edit-Distance (M-ED) algorithm and experiments to demonstrate the advantage of including unforeseen events during sequence matching. 
\item Modeling the Person Authentication using Trace Histories (PATH) problem as a variation of the person authentication using location histories \cite{PATH_MahbubChellappa2016}.
\end{itemize}

The paper is organized as follows. In Section \ref{sec:RelatedWorks}, background and related works on this topic are discussed. In Section \ref{sec:ProblemFormulation}, the approach is explained in detail along with associated challenges and possible solutions. The impact of unknown application and unforeseen events is investigated in Section \ref{sec:RoleOfUN}, and several methods for handling the active authentication problem effectively are described in Section \ref{sec:VerificationMethods}. Finally, a detailed analysis on the application usage data, experimental results and discussions are presented in Section \ref{sec:ExperimentalResults}, followed by conclusions and suggestions for future work in Section \ref{sec:conclusion}.

\section{Related Work}\label{sec:RelatedWorks}
In this section, some of the most recent published literature on active authentication and the utilization of application usage data are reviewed. The section also discusses the exploitation of two active authentication datasets (UMDAA-02 \cite{AA02_MahbubChellappa_BTAS2016} and Securacy \cite{securacy:2015}) for this research work.



\subsection{Active/Continuous Authentication of Smartphones}
Active, continuous or implicit authentication are different terminologies for the same authentication approach in which the rightful user of mobile devices is authenticated throughout the entire session of usage \cite{DemystifyingAA_Gupta2018}\cite{VMP_SPM_AA_2016}\cite{AA02_MahbubChellappa_BTAS2016}. In recent years, active authentication research has gained a lot of attention because of the increased security risks and complexity of password, token-based, multi-factor and other explicit authentication systems \cite{VMP_SPM_AA_2016}. In active authentication, the wide range of sensor data available on the mobile devices are utilized to learn one or more templates for the legitimate user during a training session. The templates are used in the background to continuously authenticate the user during regular usage and based on the amount of deviation from the templates the device itself starts restricting access to phone applications and utilities starting from the most sensitive ones \cite{AA02_MahbubChellappa_BTAS2016}. Most popular modalities for active authentication are front camera face images \cite{AAAttr_SAMANGOUEI2017_IVC}\cite{AA_Hadid}\cite{AA_Fathy}, touch screen gesture data\cite{Touchalytics}\cite{TouchAA_Feng}\cite{Heng_FG2015_Fusion}, accelerometer and gyroscope data \cite{MMC_Showme_Gambs}\cite{AccDataAuth_Primo2014}\cite{Neverova2016}, location data \cite{PATH_MahbubChellappa2016} etc. Suitability of different behavioral biometric such as touch and keystroke dynamics, phone pick-up patterns, gait dynamic, and patterns from location trace history have been explored for active authentication\cite{BehavioBiometric_MAHFOUZ2017}\cite{PhPickUp_Lee2017}\cite{PATH_MahbubChellappa2016}. Combinations of multiple biometric have been demonstrated to produce robust authentication on real-life data\footnote{\url{http://www.biometricupdate.com/201506/atap-division-head-previews-behavioral-biometrics-system-at-google-io}}.

\subsection{Prior Research on Application-Usage Data}
In recent years, there has been a lot of focus on predicting individual and community-wise application usage patterns \cite{appUsageSurvey_2017}. For example, in \cite{DiverseUsageBehaviorOfApps_Xu2011}, the authors investigate the ratio of local and global applications in the top usage list, the traffic pattern for different application categories, likelihood of co-occurrence of two different applications and such other patterns in usage.  In this work, the authors identify traffic from distinct applications using HTTP signatures. On the other hand, in \cite{InAppAd_Alok2013} the authors use mobile in-app advertisements to identify the applications in network traces. Using the ad flow data, the authors tried to analyze the usage behavior of different types of applications. In \cite{PredictUsagePattern_Xu2013}, the authors analyzed the application-usage logs of over $4,000$ smartphone users worldwide to develop an app-usage prediction model that leverages user preferences, historical usage patterns, activities and shared aggregate patterns of application behavior.

From the authentication front, in \cite{AppCentricAuth_Khan2014}, the authors proposed an application centric decision approach for active or implicit authentication in which applications are used as context to decide what modalities to use to authenticate a user and when to do it. Application usage data has also been used to generate scores for user authentication in \cite{AA_StylometryAppWeb_Friedman}. The authors only considered the frequency of occurrence of an application in the training set to determine the likelihood of being a particular user, missing the temporal variation in the the usage pattern. 

An interesting use-case of application-usage data is presented in \cite{ShowAppFindFriend_Shema2017}. The authors used a large-scale annotated application-usage dataset to build a predictor that can estimate where a person is (\textit{e.g.}, at home or office) and if he/she is with a close friend or a family member. 
In \cite{BatteryLifefromUsage_Houran2018}, the authors used application usage traces along with system status and sensor indicators to predict the battery life of the phones using machine learning techniques.  

\subsection{Datasets on Application Usage}
Even though there have been diverse research approaches that need application-usage data, there is a scarcity of publicly available datasets. Also, many of the application-usage datasets have limited number of applications or are not unbounded real-life usage data, but instead contain data generated under supervision or by following certain instructions. In this work, all the experiments are performed on two well-known large scale public datasets suitable for investigating the active authentication problem, namely, the application-usage data of University of Maryland Active Authentication Dataset-02 (UMDAA-02)\footnote{Available at \url{https://umdaa02.github.io/}} \cite{umdaa02:2016} and the Securacy \footnote{Available at \url{http://ubicomp.oulu.fi/securacy-understanding-mobile-privacy-and-security-concerns/}} \cite{securacy:2015} dataset from the Center of Ubiquitous Computing, University of Oulu.

\subsubsection{UMDAA-02 Application-Usage Dataset}

\begin{table}
\centering
\caption{General information on application-usage data available in the UMDAA-02 dataset.}
\begin{tabular}{p{6.0cm} c}
\hline
No. of Subjects with $\geq 500$ training samples and $\geq 200$ test samples for sampling rate of $1/30s^{-1}$	(Train/Test)									&$32/26$\\ 
\hline
Avg. No. of Sessions/User with App-Usage Data  of the $26$ selected subjects (train/test)  	&$\sim 582/\sim 197$\\
\hline
Train/Test split for the experiment							&$70\% / 30\%$ \\
\hline
Total Number of Unique Applications Used by the $26$ selected subjects (train/test) & $119/67$ \\
\hline
Average Number of Samples Per User for the $26$ selected subjects (train/test)	&$\sim 4307/\sim 1399$ \\
\hline
\end{tabular}
\label{AppDataUMDAA02}
\vskip -5pt
\end{table}

\begin{figure*}[t]
\centering
\subfigure[]{\includegraphics[width = 0.45\textwidth]{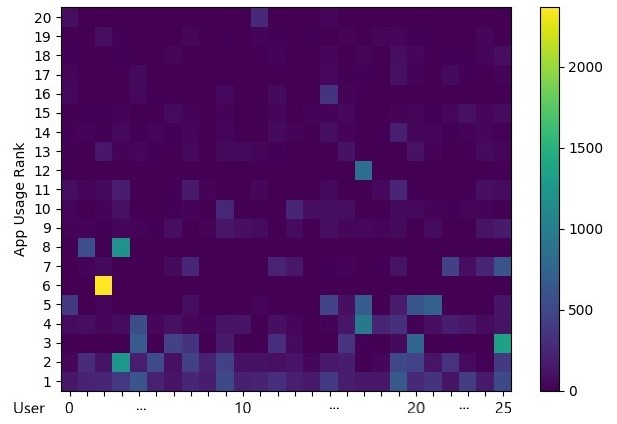}}~
\subfigure[]{\includegraphics[width = 0.45\textwidth]{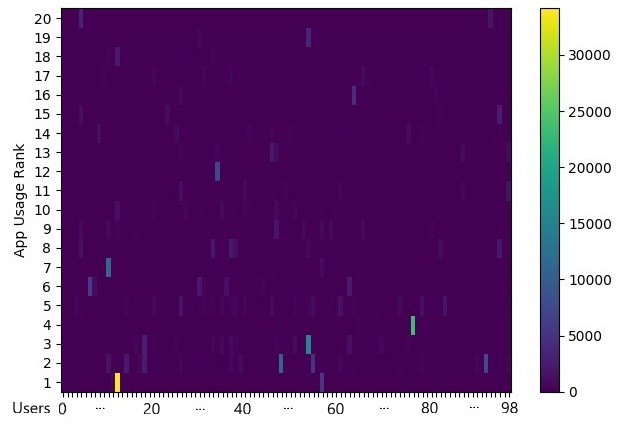}}~
\caption{Similarity matrix depicting top $20$ application-usage rate among users in the training set of the (a) UMDAA-02 dataset and, (b) Securacy dataset.}
\label{TopAppHistogram_bothDataset}
\end{figure*}

The UMDAA-02 dataset is specifically designed for evaluating active authentication systems in the wild. The dataset consists of 141.14 GB of smartphone sensor data collected from $45$ volunteers who were using Nexus 5 phones in their regular daily activities over a period of two months. The data collection application ran completely in the background and the collected data includes the front-facing camera, touchscreen, gyroscope, accelerometer, magnetometer, light sensor, GPS, Bluetooth, WiFi, proximity sensor, temperature sensor and pressure sensor among with the timing of screen unlock and lock events, start and end timestamps of calls and currently running foreground application, etc. The application usage data from $45$ users is summarized in Table \ref{AppDataUMDAA02}. However, not all the users have adequate amount of usage data. For all the experiments in this paper, a total of $26$ users are used who has more that $500$ training samples and more than $200$ test samples for any sampling rate between $1/5s^{-1}$ to $1/30s^{-1}$. The usage statistics for the top 20 applications for the selected $26$ subjects is presented in Table~\ref{tab:TopAppInfoUMDAA02}. The usage rate for the top $20$ applications for each user is shown in fig. \ref{TopAppHistogram_bothDataset}(a). From the table and the figure, it is readily seen that the applications ranked $6$th, $8$th, $12$th and $20$th are in the top list because of excessive usage by very few users, where as, the remaining applications are genuinely popular among the users.

\begin{table}
  \centering
  \caption{App-usage statistics for the top $20$ apps for the $26$ selected users of the UMDAA-02 dataset.}
\begin{tabular}{p{0.5cm} p{3.5cm} p{0.5cm} p{1cm} p{1cm}}
\hline
Rank 	& App Name 	& No. of Users	& Per User Usage 	& Overall Usage 	\\
\hline
    1     & com.google.android. googlequicksearchbox & 26    & 283.27 & 283.27 \\
\hline
    2     & com.android.dialer & 25    & 255.24 & 245.42 \\
\hline
    3     & com.whatsapp & 15    & 303.6 & 175.15 \\
\hline
    4     & com.android.chrome & 26    & 141.42 & 141.42 \\
\hline
    5     & com.facebook.katana & 11    & 308.18 & 130.38 \\
\hline
    6     & com.nextwave.wcc2 & 1     & 2366  & 91 \\
\hline
    7     & com.google.android.youtube & 16    & 144.38 & 88.85 \\
\hline
    8     & com.ea.game.pvzfree & 2     & 872.5 & 67.12 \\
\hline
    9     & com.google.android.gm & 24    & 51.04 & 47.12 \\
\hline
    10    & com.android.mms & 22    & 52.09 & 44.08 \\
\hline
    11    & com.google.android.talk & 18    & 62.28 & 43.12 \\
\hline
    12    & com.andrewshu.android.reddit & 1     & 842   & 32.38 \\
\hline
    13    & com.nextbus.mobile & 19    & 41.89 & 30.62 \\
\hline
    14    & com.google.android.apps.docs & 24    & 33    & 30.46 \\
\hline
    15    & com.android.settings & 24    & 27.71 & 25.58 \\
\hline
    16    & com.google.android.apps.maps & 14    & 44    & 23.69 \\
\hline
    17    & com.android.camera2 & 22    & 20.5  & 17.35 \\
\hline
    18    & com.google.android.gallery3d & 17    & 24.94 & 16.31 \\
\hline
    19    & com.android.vending & 21    & 20.1  & 16.23 \\
\hline
    20    & com.viber.voip & 5     & 74.6  & 14.35 \\
\hline
    \end{tabular}%
  \label{tab:TopAppInfoUMDAA02}%
\end{table}%

\subsubsection{Securacy Application Usage Dataset}

The Securacy dataset was originally created within the context of exploring the privacy and security concerns of a smartphone user by analyzing the location of servers that different applications use and whether secure network connections are used. For a period of approximately six months, the data was collected from 218 anonymous participants who installed the data collection application from the Google Play store. The collected data, 679.90 GB, includes the currently running foreground application, installed, removed or updated applications, application server connections and device location, etc. Out of the $218$ users of the original Securacy dataset, $99$ are used for this experiment based on the limits on training and test observations as mentioned for the UMDAA-02 dataset. The application usage data for the $99$ subjects in the Securacy dataset are summarized in Table~\ref{AppDataSecuracy} and the corresponding usage statistics for the top 20 applications are presented in Table~\ref{tab:TopAppInfoSecuracy02}. The usage rate for the top $20$ applications for each user are show in fig. \ref{TopAppHistogram_bothDataset}(b). Note that the top applications ranked $1$st, $2$nd and $4$th in Table~\ref{tab:TopAppInfoSecuracy02} are actually the same application written in Spanish, English and Finnish, respectively. Similarly, rank $12$, `Horloge' is 'Clock' in French, and therefore is the same application as rank $19$. However, these applications are shown separately here because, for the active authentication problem, even the preferred language of the user is a type of biometric metadata and can be used to discriminate between users. Also, similar to UMDAA-02 dataset usage statistics, there are several applications in the top $20$ rank that were actually used by only a few users very frequently (ranked 1, 4, 9, 12, 16). For this dataset, this phenomenon can be attributed to language difference as well because if the language difference were nullified, then rank 1, 2, 4 will collapse at rank 1 and rank 12 and 19 will collapse at 12 - thereby removing three applications from the list (rank 1, 4 and 12) that has very few users. For the user verification research presented here, the language variation is kept unaltered in order to retain the naturalness of the dataset and the algorithms are expected to learn to discriminate between users based on the language as well as on usage pattern. 
%
%

\begin{table}
\centering
\caption{General information on application-usage data available in the Securacy dataset.}
\begin{tabular}{p{6.0cm} c}
\hline
No. of Subjects with $\geq 500$ training samples and $\geq 200$ test samples for sampling rate of $1/30s^{-1}$ (Train/Test)									&$201/99$\\ 
\hline
Avg. No. of Sessions/User with App-Usage Data  of the $26$ selected subjects (train/test)  	&$\sim 119/\sim 96$\\
\hline
Train/Test split for the experiment							&$70\% / 30\%$ \\
\hline
Total Number of Unique Applications Used by the $26$ selected subjects (train/test) & $1340/554$ \\
\hline
Average Number of Samples Per User for the $26$ selected subjects (train/test)	&$\sim 2235/\sim 1745$ \\
\hline
\end{tabular}
\label{AppDataSecuracy}
\vskip -5pt
\end{table}

\begin{table}[t]
  \centering
  \caption{App-usage statistics for the top $20$ apps for the $99$ selected users of the Securacy Dataset.}
\begin{tabular}{p{0.5cm} p{3.5cm} p{0.5cm} p{1cm} p{1cm}}
\hline
Rank 	& App Name 	& No. of Users	& Per User Usage 	& Overall Usage 	\\
\hline
    1     & Sistema Android & 4     & 9972.25 & 402.92 \\
%
%
%
%
%
%
\hline
    2     & Android System & 80    & 480.44 & 388.23 \\
\hline
    3     & com.android.keyguard & 34    & 802.79 & 275.71 \\
\hline
    4     & Android-jrjestelm & 5     & 4820.8 & 243.47 \\
\hline
    5     & System UI & 80    & 242   & 195.56 \\
\hline
    6     & Nova Launcher & 19    & 794.79 & 152.54 \\
\hline
    7     & Maps  & 38    & 363.08 & 139.36 \\
\hline
    8     & Google Search & 53    & 214.3 & 114.73 \\
\hline
    9     & Launcher & 12    & 650   & 78.79 \\
\hline
    10    & Chrome & 60    & 128.2 & 77.7 \\
\hline
    11    & Facebook & 49    & 154.53 & 76.48 \\
\hline
    12    & Horloge & 1     & 7328  & 74.02 \\
\hline
    13    & YouTube & 49    & 144.94 & 71.74 \\
\hline
    14    & TouchWiz home & 20    & 348.3 & 70.36 \\
\hline
    15    & Securacy & 84    & 75.39 & 63.97 \\
\hline
    16    & Internet & 16    & 371.25 & 60 \\
\hline
    17    & WhatsApp & 37    & 154.62 & 57.79 \\
\hline
    18    & Google Play Store & 72    & 71.83 & 52.24 \\
\hline
    19    & Clock & 44    & 113.89 & 50.62 \\
\hline
    20    & Package installer & 36    & 138.69 & 50.43 \\
\hline
    \end{tabular}%
  \label{tab:TopAppInfoSecuracy02}%
\end{table}%

\section{Problem Formulation}\label{sec:ProblemFormulation}
The application usage data from smartphones coupled with the timing information can be used to determine the exact day time and duration of using any application. It is assumed here that there might be certain pattern in the usage of different applications at different time of the day or during weekdays and weekends. Hence, a state-space model can be intuitively considered for modeling the pattern of application usage for a particular user. Models for different users are assumed to be different because of the difference in lifestyle of each individual. Therefore, the state-space model of a user can effectively be considered as a model for the pattern of life of that user and can be used to differentiate the user from others. There are however several challenges to this approach towards solving the authentication problem using application usage:
\begin{itemize}
\item Forming observation states from the application data and corresponding timing information.
\item Training a state-space model in a way that it can handle unforeseen observations during testing.
\item Generating verification scores from sequential observation data.
\end{itemize}
Each of these challenges and the proposed solutions are discussed here. 

\subsection{Application Names to Observation States}
Incorporating the temporal information with the application name is a challenge because the user can use an application at any time, and therefore power set of all applications and all probable time is intractable even if we sample at a relatively high frequency. For example, if there are $N$ number of applications and if we sample every $5$ minutes, then there would be $480$ unique time stamps in a day and $3360$ timestamps in a week. This would mean a total of $3360\times N$ observation states for the applications in a week. However, for a single application, most of these observation states will either not occur or occur very infrequently in the training set. Hence, training a reliable state-space models with this sparsely occurring observation states will be difficult. 

\begin{figure*}[t]
\centering
\subfigure[]{\includegraphics[width = 0.49\textwidth]{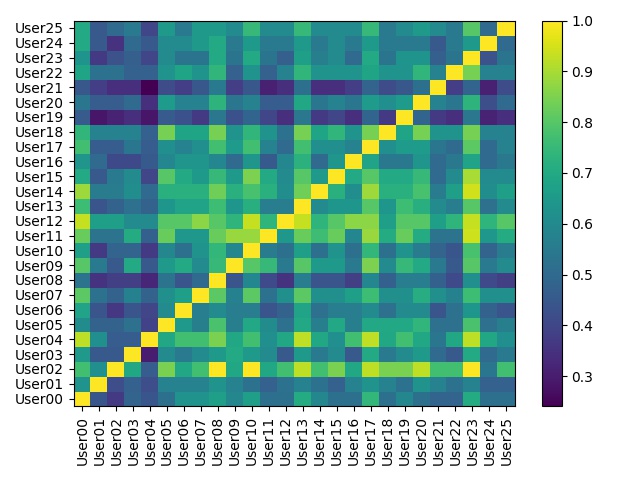}}~
\subfigure[]{\includegraphics[width = 0.49\textwidth]{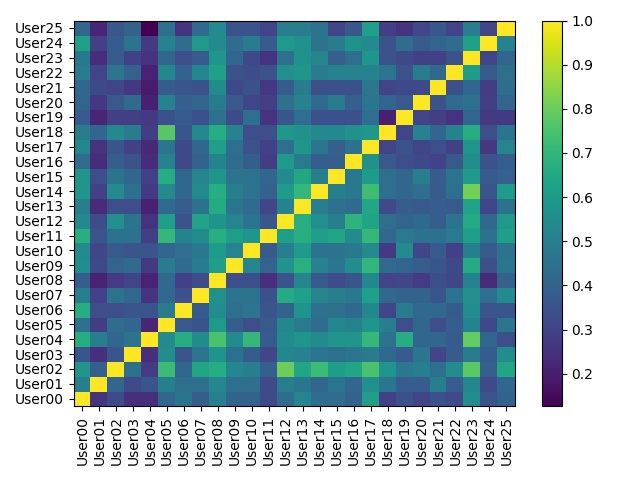}}~

\subfigure[]{\includegraphics[width = 0.49\textwidth]{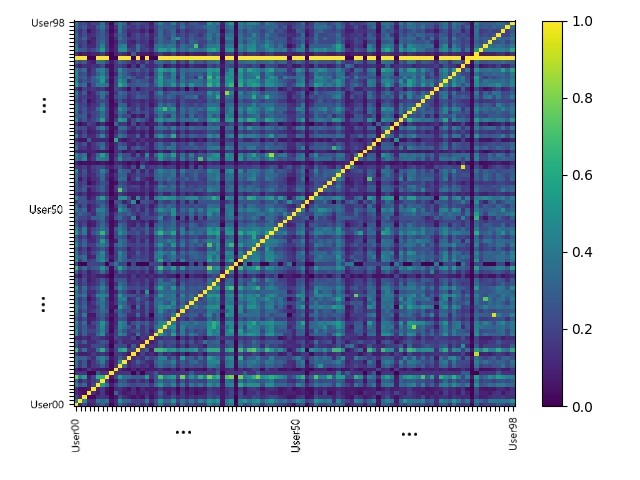}}~
\subfigure[]{\includegraphics[width = 0.49\textwidth]{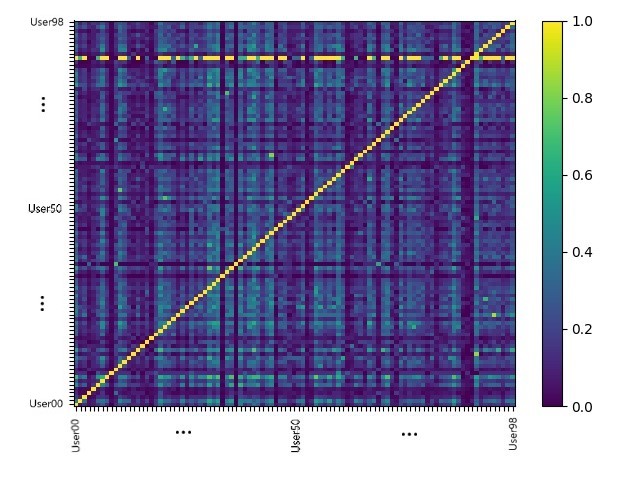}}~
\caption{Similarity matrix depicting (a) application name overlap, and (b) observations overlap for the training set of the UMDAA-02 dataset. Similarly, (c) and (d) depicts the application overlap and observations overlap for the training set of the Securacy dataset}
\label{AppUsageSimilarityMatrix_UMDAA02_Securacy}
\end{figure*}

In this regard, the time-zone and weekday/weekend flag idea are adopted from \cite{PATH_MahbubChellappa2016}. By dividing the day into three distinct time zones (TZs), namely, $TZ_1$ (12:01 am to 8:00 am), $TZ_2$ (8:01 am to 4:00 pm) and $TZ_3$ (4:01 pm to 12:00 pm), and denoting weekday/weekend with a flag $W(t)\in{W_D, W_E} \forall t$, respectively, the total number of possible observation states is kept limited to $6N$. The functions $TZ(t)$ and $W(t)$ maps any time $t$ into one of the corresponding timezone and weekday/weekend, respectively. The impact of converting application tags into observations on verifying the users of the UMDAA-02 app-usage data and the Securacy datasets can be visualized from Figs. \ref{AppUsageSimilarityMatrix_UMDAA02_Securacy}(a)-(b) and \ref{AppUsageSimilarityMatrix_UMDAA02_Securacy}(c)-(d), respectively. The similarity matrix in Figs. \ref{AppUsageSimilarityMatrix_UMDAA02_Securacy}(a) depicts the percentage of common applications between two users in UMDAA-02 training dataset, whereas, the similarity matrix in Fig. \ref{AppUsageSimilarityMatrix_UMDAA02_Securacy}(b) depicts the percentage of common observations between any two users on the same dataset. It is clear that the similarity of observations between two different users is less than the similarity of applications. The effect is less visible on the Securacy dataset (Figs. \ref{AppUsageSimilarityMatrix_UMDAA02_Securacy}(c)-(d))because the subjects came from a diverse population than the subjects of the UMDAA-02 dataset. Hence, the similarity of applications is less pronounced, yet, the differences between application similarity and observation similarity are still present.

\subsection{Taking Unknown Applications into Account}
Now, in order to handle unknown applications that might be present in the test set, an additional application name $U$ is considered. The $U$ application adds $6$ observation states when combined with TZs and $W$. Note that in the training set there is no probability of having any $U$ application, and all the observations with $U$ are assigned a very small prior probability ($10e-20$) when state-space models are trained. Also, it is ensured for state-space models that the emission probability for the states with $U$ application does not go to zero, in order to prevent zero probability score during testing when unknown applications are encountered. If the total number of unique applications used by user $X$ in the training set is $A_x$, then any application $\alpha_y$ of the test user $Y$ in the test set $\bar{A}_y$ will be denoted as $U$ if $\alpha_y \notin A_x$. In \cite{PATH_MahbubChellappa2016}, the authors addressed similar issues for geo-location data by considering even more additional states such as nearby unknowns. However, proximity is a vague concept for application data and therefore only $U$ is considered here. Note that, any observation with an unknown application is unforeseen by default, but an unforeseen observation with some other application name is not unknown.  

Note that, apart from $U$, unforeseen observations might be present in the test set. For example, in the training set an application $\alpha_x$  might only occur in weekdays at timezones $TZ_1$ and $TZ_2$ while the same application might be used in the test set at time zone $TZ_3$ on a weekday. In that case, the test observation $(\alpha_x, TZ_3, W_D)$ would be unforeseen in the training set. For state space models, this problem is handled by generating all possible combinations applications, time zone and day flag and use them to construct the model. If one such observation is not present in the training set, it is assigned non-zero prior and emission probabilities to ensure that they do not bring down the probability of a test sequence to zero. 

\subsection{Handling Uncertainty}
Now that unknown applications and unforeseen observation states are addressed, we tackle the creation of observation states via binning of time-stamped data. In most cases, the data collection is done in sessions, where a session starts with unlocking the phone and stops when the phone is locked again. Even if this is not the case, there can be very long idle times between consecutive usage of a phone, during which, authentication is a redundant operation and no application is running in the foreground \cite{vanBerkel:2016:SAS:2858036.2858348}. Hence, there can be a big gap between the start-time for an application and the stop time of the previous application in the data log. This time gap might be as small as several seconds to as big as several days even for a user who owns a smartphone for regular use \cite{Church:2015:UCM:2785830.2785891}. The sparsity introduced by this time gap is handled in two ways. At the beginning of each session (unlocking of the phone) a dummy observation state $\Psi$ is introduced. The state-space model is expected to learn that $\Psi$ is a time gap which might or might not cause a change in the time zone. For example, the last used application might be in $TZ_1$ before the closing of a session. Then the next session may occur in either $TZ_1$ or $TZ_2$ or $TZ_3$ of the same day. If the next session is in the next day or if the day changes within a running session, then an additional flag $\Delta$ is introduced which denotes the transition into next day. The time zone and weekday/weekend flags are ignored for observations $\Psi$ and $\Delta$.
%
%

So, taking the six probable observations for $U$ and the $\Psi$ and $\Delta$ observations into consideration, the total number of possible observation states for user $X$ would be $6N+6U+\Psi+\Delta$. 

\begin{figure*}[t]
\centering
\includegraphics[width = 0.99\textwidth]{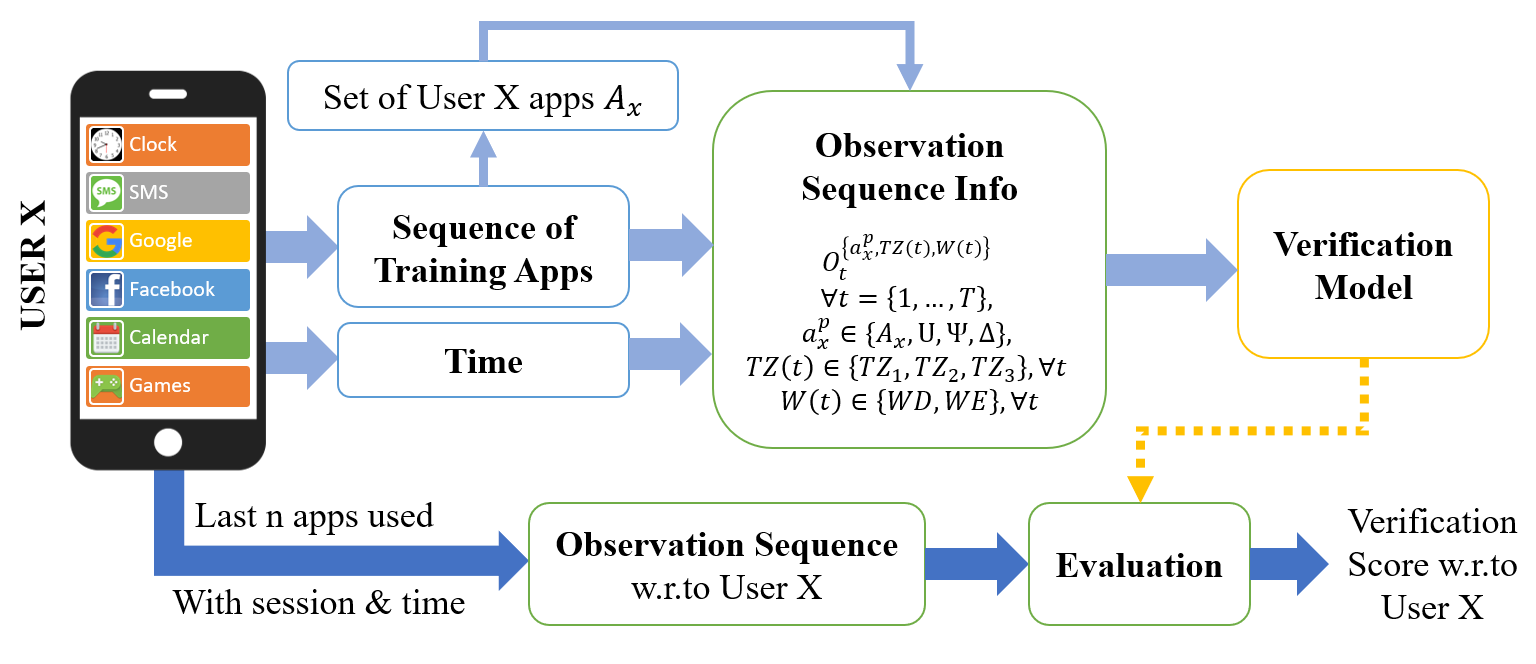}
\caption{Overview of an application-usage-based user verification system for mobile devices.}
\label{SystemDiagram}
\end{figure*}

\subsection{System Overview}
A diagram depicting an application-usage-based user verification system is shown in Fig. \ref{SystemDiagram}. Once the observation sequence is extracted, a verification model can be trained based on the patterns in the sequence. The verification model can be a state space model, a string matching approach or even a recurrent neural network, depending on data availability and need. For state-space models, once training for a user is done, the model can be used to generate scores for last $n$ test observation sequences created using the same protocol that was used during the training phase. The score can be thresholded to obtain the verification decision. For more simpler methods such as sequence matching, unknown applications and unforeseen observations are difficult to handle. For the authentication problem, the unknown and unforeseen play key roles, described in the next section.

\section{The Role of Unknown application and Unforeseen Observations in User Verification}\label{sec:RoleOfUN}
\subsection{Statistics of unknown applications in the test data}
\begin{figure*}[t]
\centering
\subfigure[]{\includegraphics[width = 0.8\textwidth]{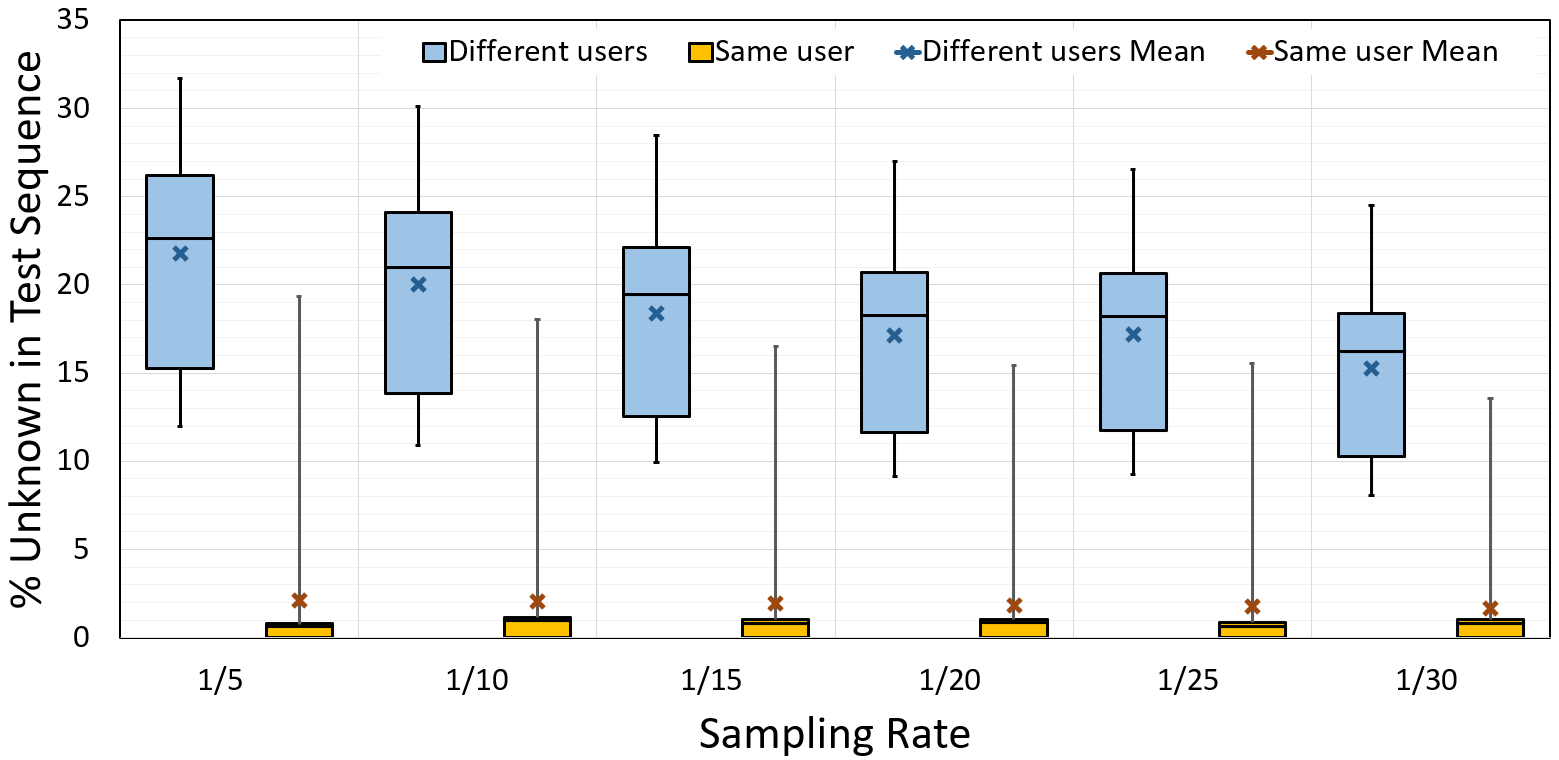}}~

\subfigure[]{\includegraphics[width = 0.8\textwidth]{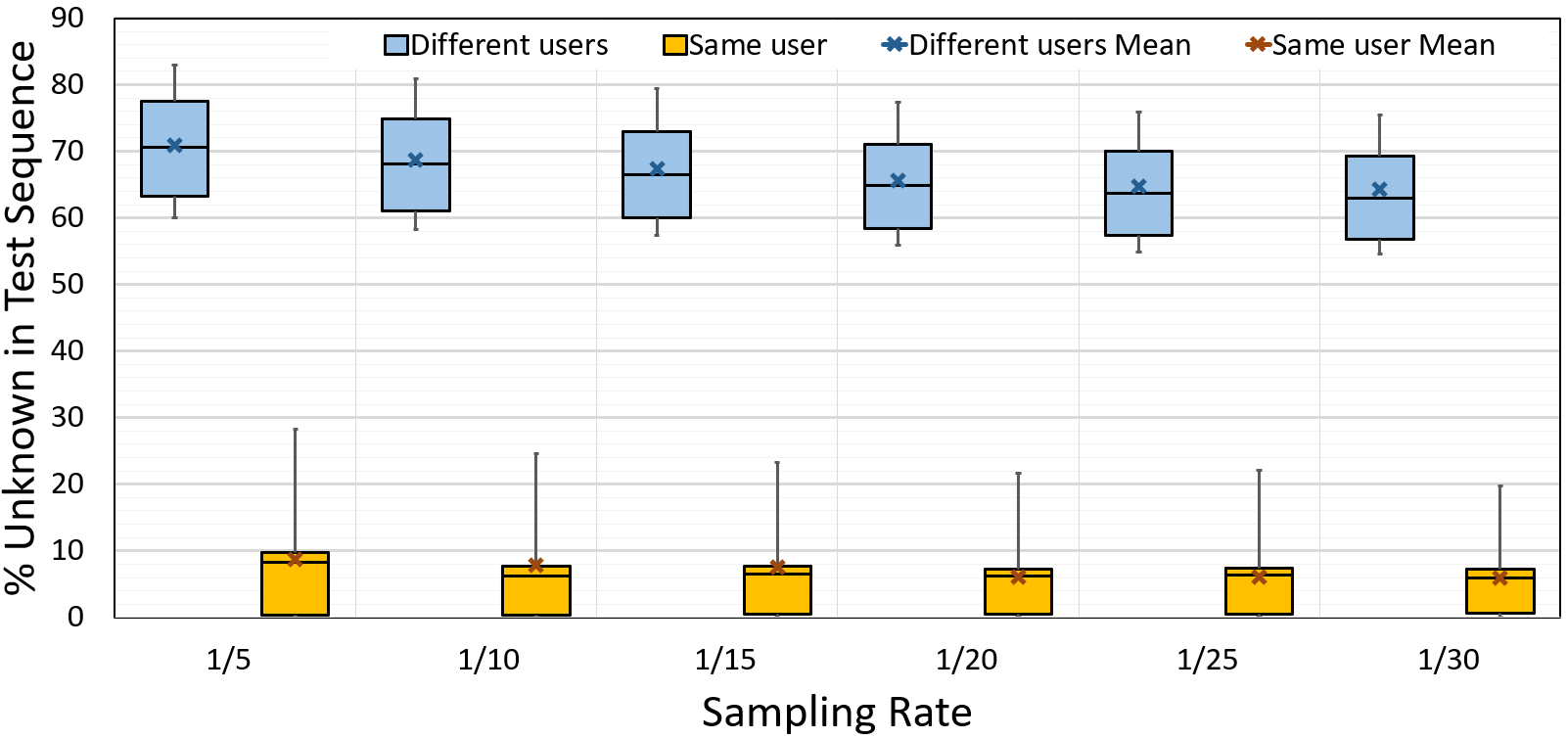}}~
\caption{Boxplots depicting the percentage of unknown application in test data for (a) UMDAA-02 dataset, and (b) Securacy dataset, for different sampling rates. Note that the average percentage of unknown applications used by the the different user is much bigger than that for same user on both datasets.}
\label{Unk_Stat}
\end{figure*}

If an application is present in the test set but not encountered in the training set, the application is denoted with $U$ as unknown application in the proposed formulation. Intuitively, the prevalence of $U$ will be much higher if the test set comes from a different user or from an intruder of the phone, while for the legitimate user the test set will have fewer unknown applications. This intuition is verified on the application usage data from both UMDAA-02 and Securacy datasets, as can be seen from the box plots in Fig. \ref{Unk_Stat}. 

Note that the gap between the whisker plots for same user and different users is larger for the Securacy dataset in comparison to UMDAA-02 dataset. Securacy is a larger dataset with more users, more data per user and more variation in user demographies compared to UMDAA-02 in which the subjects were from a narrow age range and were all affiliated with the same institution. Hence, it shows that among the general population, even the selection of applications varies widely between users. 

\subsection{Impacts of binary decision based on unforeseen events}
Two simple experiments with unknown applications $U$ and unforeseen observations are performed on the UMDAA-02 and Securacy datasets to evaluate their role in user verification. The observations for each user are chronologically sorted and the earliest $70\%$ observations are considered for training and the rest for testing. Now, for any user $i$ in the training set, a sequence of training observations $S_i^{tr}$ is obtained along with the set of unique applications $A_i$. Now, each test sequence of a user is compared with the training sequence and application lists of the training subjects and different binary hard decision rules are applied in two experiments. In the first experiment, the binary decision rule is based on occurrence of an application in the test set that is not present in the training set. In the second experiment, the decision is taken based on the occurrence of an unforeseen observation in the test set. In both cases, if there is even a single occurrence of an unknown application or an unforeseen observation, then the match score is set to $0.0$, otherwise it is set to $1.0$. The matching algorithms for the two experiments are shown in (\ref{algo:BinUnk}) and (\ref{algo:BinUnfor}), respectively. The data sampling rates for both these experiments were set to $1/30$ per second, which resulted in $\sim 16863$ training-test sequence pairs for the UMDAA-02 application-usage dataset and $\sim 846331$ training-test sequence pairs for the Securacy dataset. The number of users with adequate training and test data is $26$ in UMDAA-02  and $99$ in Securacy, leading to an average of $\sim 647$ and $\sim 8549$ pairs per user, respectively. 
%
%

\begin{algorithm}[h]
\caption{Binary Decision Rule based on Unknown Applications}\label{algo:BinUnk}
\begin{algorithmic}
\Procedure{BinUnk}{$A_i$, $S^{te}_j$}\Comment{List of unique applications of user $i$ ($A_i$), n-last Test Sequence Vector of user $j$ ($S^{te}_j$)}
\For{$v^{te} \in S^{te}_{j}$}\Comment{Loop through all test observations}
	\State $a^{te} \gets v^{te}[0]$\Comment{Get the application name from the test observation}
	\If {$a^{te} \notin A_{i}$}
        \State \textbf{return} $0.0$ \Comment{Return score 0.0 if any unknown application is encountered}
 		\EndIf
	\EndFor
\State \textbf{return} $1.0$ \Comment{Return score 1.0 if no unknown application in test sequence}   
\EndProcedure
\end{algorithmic}
\end{algorithm}

\begin{algorithm}[h]
\caption{Binary Decision Rule based on Unforeseen Observations}\label{algo:BinUnfor}
\begin{algorithmic}
\Procedure{BinUnfore}{$S^{tr}_i$, $S^{te}_j$}\Comment{Sequence of training observations for user $i$ ($S^{tr}_i$), n-last Test Sequence Vector of user $j$ ($S^{te}_j$)}
\For{$v^{te} \in S^{te}_{j}$}\Comment{Loop through all test observations}
	\If {$v^{te} \notin S^{tr}_i$}
        \State \textbf{return} $0.0$ \Comment{Return score 0.0 if any unforeseen observation is encountered}
 		\EndIf
	\EndFor
\State \textbf{return} $1.0$ \Comment{Return score 1.0 if no unforeseen observation in test sequence}   
\EndProcedure
\end{algorithmic}
\end{algorithm}

\begin{figure*}[t]
\centering
\includegraphics[width = 0.9\textwidth]{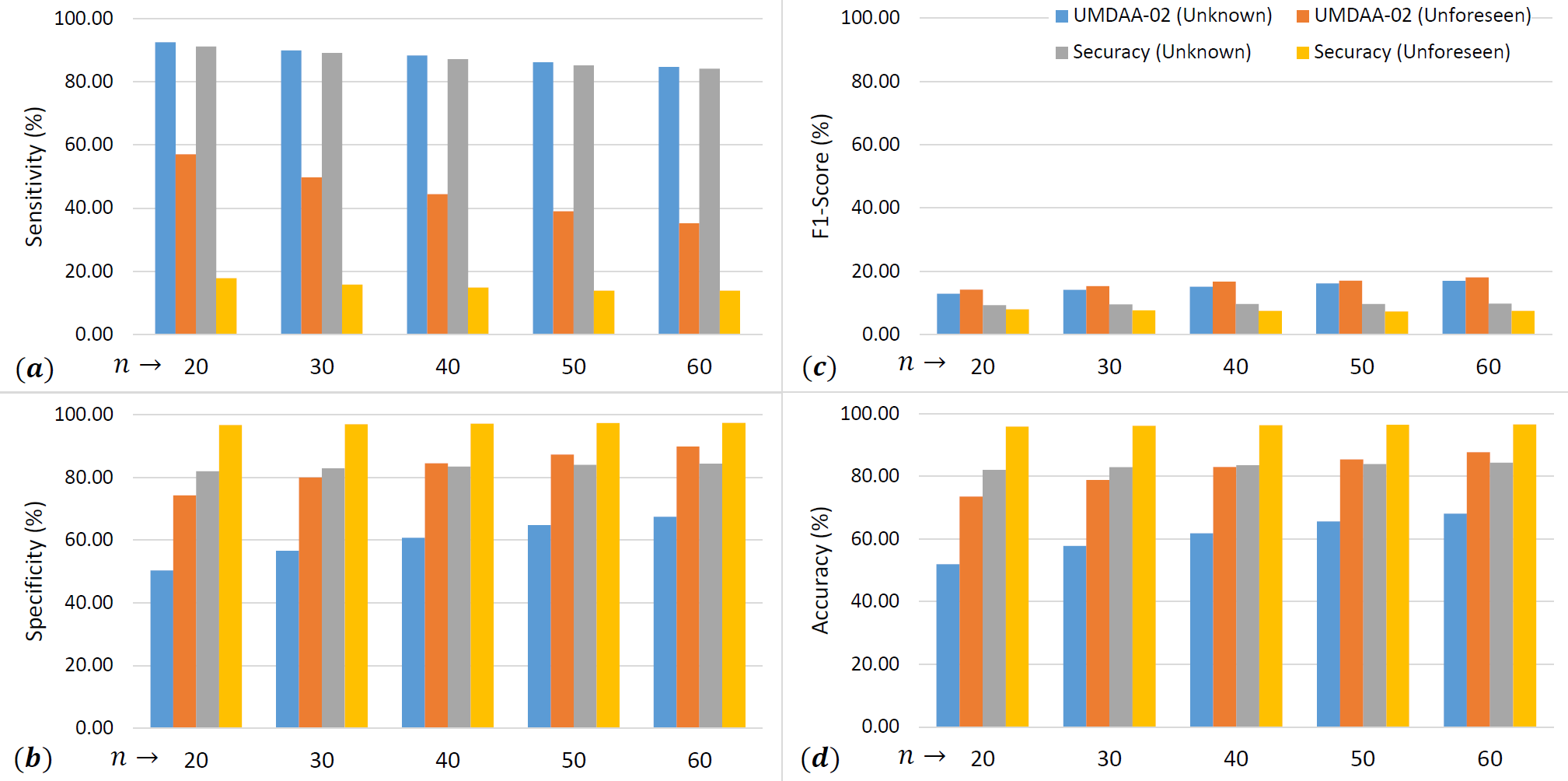}
\caption{(a) Sensitivity, (b) Specificity, (c) F1-Score, and (d) Accuracy (in $\%$) obtained by varying sequence length $n$ for Securacy and UMDAA-02 application-usage data for using the Binary Hard Decision rule based on unknown applications and unforeseen observations.}
\label{BinaryHardDecisionOnUnk}
\end{figure*}

Results for several evaluation metrics namely, sensitivity, specificity, F1-score and accuracy - all in percentage, obtained through the two experiments on the two datasets are shown in Fig. \ref{BinaryHardDecisionOnUnk}(a)-(d). The definition of these metrics are as follows:
\begin{eqnarray}
Sensitivity&=&\frac{TP}{TP+FN}\times 100\%\\
Specificity&=&\frac{TN}{TN+FP}\times 100\%\\
Accuracy&=&\frac{TP+TN}{TP+FP+TN+FN}\times 100\%\\
F1-Score&=&\frac{2TP}{2TP+FP+FN}\times 100\%
\end{eqnarray}
where, $TP$, $FP$ and $FN$ are the numbers of true positive,  false positive and false negative detections, respectively. High Sensitivity implies smaller number of false-negatives, while high Specificity implies less false-positives. Accuracy over $50\%$ denotes that the true values outweighs the false predictions. Finally,  F1-Score implies better overall precision and recall.

Fig. \ref{BinaryHardDecisionOnUnk} gives the following interesting insights about the impact of the unknown applications and unforeseen observations on the performance metrics for the two datasets. 
\begin{itemize}
\item With increasing sequence length $n$, the specificity is increasing gradually for all the cases, while sensitivity is decreasing. The decrease in sensitivity is probably due to the fact that the probability of having an unknown application in the sequence increases with increasing sequence size, thereby increasing the chances for false negatives. On the other hand, with increasing $n$ more sequences are denoted as negatives, which in effect reduces the number of false positives and therefore increases specificity.
\item The sensitivity drops drastically when unforeseen observations are used instead of unknown applications as decision criteria. This is understandable, since the number of false negatives increases rapidly when all sequences with at least on unforeseen observations are marked as data from a different user.
\item The number of false positives decreases when unforeseen observations are considered for decision instead of unknowns. This leads to a jump in specificity for a fixed $n$. In general, the specificity is much higher for Securacy dataset in comparison to UMDAA-02. This proves that there are more unknown applications and unforeseen applications in Securacy when comparing a user with others. Securacy being a more diverse and larger dataset has wider variation of information, which leads to this phenomenon. Here, the training data for each user is longer, meaning that they are much closer representation of real life and therefore, an unknown application or unforeseen observation is actually a different user's data in most cases. 
\item Higher sensitivity, however, does not mean that for real life data a simple binary classifier based on unforeseen observations is reasonably good. The F1-Score is very low for both datasets, which means either precision or recall or both of therm are very low. Since in the active authentication using application usage, the number of positive pairs is largely outweighed by the number of negative pairs, it can be assumed that $FP>>FN$ and $TN>>TP$. Since Precision=$\frac{TP}{TP+FP}$ and Recall=$\frac{TP}{TP+FN}$, that means, Recall$>$Precision. With increasing $n$, $FN$ increases, while $FP$ decreases, leading to reduction in recall and increase in precision. However, given the fact that the F1-Score does not improve much with increasing $n$, it can be assumed that Recall reduces steeply while Precision does not improve much. 
\item Irrespective of deciding with unknown or unforeseen, the accuracy is always lower for the UMDAA-02 dataset in comparison to Securacy dataset. Even though the application-usage information in Securacy is much larger than UMDAA-02, probably due to the high demographical similarity among the subjects of UMDAA-02, the binary hard measure performs poorly in comparison to Securacy. In practice, there could not be any assumption made about the demographic similarity or dissimilarity of an user and an intruder - hence, using neither unknown applications nor unforeseen observations as a hard decision metric cannot be a practical solution to the active authentication problem. 
\item The experiment once again proves that 'accuracy' is not a good performance metric when the number of samples between classes is severely biased. In this example, the average percentage of positive pairs in the dataset is $\sim 3.85\%$  on UMDAA-02 dataset and $\sim 1.01\%$ in the Securacy dataset. Being an open set problem, the task is to deal with heavily biased data towards negative samples and better performance measures in this regard would be receiver operating characteristic (ROC) curves and equal error rates (EER) instead of accuracy.
\end{itemize}

\begin{algorithm*}
\caption{Pseudocode for the modified edit-Distance algorithm.}\label{algo:MEDist}
\begin{algorithmic}
\Procedure{M-ED}{$S^{tr}$, $S^{te}$}\Comment{Training observation sequence of a user ($S^{tr}$) of length $N$, $n$-last Test observation Sequence of any user ($S^{te}$), where length($S^{tr}$)$>n$.}
\State $D \gets [1, 2, \hdots, n]$
\For{$j = 0$ \textbf{to} $n-1$}
	\State $d \gets zeros[1:n]$
	\State $d[0] \gets [j+1]$
    \For{$i = 0$ \textbf{to} length($S^{tr}-1$)}
    	\If{$S^{te}[j]==S^{tr}[i]$}
        	\State $d[i+1] \gets D[i]$\Comment{Exact match, no operation needed.}
    	\Else
        	\State $A_1, T_1, W_1 \gets S^{tr}[i]$ \Comment{Extract application name, timezone and day flag from the observations.}
        	\State $A_2, T_2, W_2 \gets S^{te}[j]$ \Comment{Extract application name, timezone and day flag from the observations.}
            \State NOp $\gets 0$
        	\If{$A_1==A_2$ \textbf{ and } ($T_1==T_2$ \textbf{ or } $W_1==W_2$)}
	        	\State NOp $\gets 1$ \Comment{One substitution needed if only timezone or day does not match.}
            \ElsIf{$A_1==A_2$}
                \State NOp $\gets 2$ \Comment{Two substitution needed if neither timezone nor day are matching.}
            \Else 
	        	\State NOp $\gets 3$ \Comment{Three substitution operation for no match.}
           	\EndIf
            \State $d[i+1] \gets$ NOp$+\min(D[i], D[i+1], d[i-1])$      	
        \EndIf
        \State $D \gets d$
    \EndFor
\EndFor

\State \textbf{return} $D[n-1]$  
\EndProcedure
\end{algorithmic}
\end{algorithm*}

\subsection{Impacts of ignoring unforeseen events}

Now that the impact of unforeseen events on the authentication problem are established, a slightly more advanced sequence matching approach based on Levenshtein Distance a.k.a Edit-Distance (ED) is performed to study the impact of ignoring the unknown observations and unforeseen events. When matching sequence $s_1$ to another sequence $s_2$ of the same length, the original ED calculates the number of deletions, insertions, or substitutions required to transform $s_1$ to $s_2$. For the active authentication problem, let's assume that a test observation sequence $S^{te}$ of length $n$ is to be matched with any training observation sequence $S^{tr}$ of length $N$, where, intuitively $N>n$. Since each observation consists of an application name, timezone and day flag, when a mismatch occurs, the distance can be assumed to be different depending on the amount of match. For example, if only the application name matches, then the timezone and day flag needs to be substituted, leading to two operations. Based on this fact, the a modified algorithm for edit distance (M-ED) is presented in (\ref{algo:MEDist}). Using this algorithm, three different tests are performed on the UMDAA-02 dataset, the results for which are given in Table \ref{tab:ModEDRes}. In the first test, all test observations are included, while in the next two tests, the observations with unknown applications, and the unforeseen observations are ignored. In order to ignore the unforeseen observations, for any training sequence, each test sequence is compared to find the unforeseen observations and removed from the test sequence. For unknown applications, the corresponding observation is removed. This operation reduced the number of samples per user from $891$ to $458$ and $245$, respectively, and the number of unique application in the test data went from $61$ to $60$ and $45$. As can be seen from Table \ref{tab:ModEDRes}, the lowest EERs for any value of $n$ are obtained when all observations are considered. Ignoring both unknown applications and unforeseen observations make the verification task difficult. Also, for practical purposes, ignoring samples will cause latency in decision making, which can greatly reduce the recall of an active authentication system.

\begin{table}[t]
\centering
\caption{Performance of the M-ED algorithm in terms of EER (\%) for three types of test sequences - all observations, all except the ones with unknown applications and all without unforeseen observations. Experiment performed on the UMDAA-02 dataset with fixed sampling rate at 1/30$s^{-1}$.}
    \begin{tabular}{cccc}
    \toprule
    \multicolumn{1}{c}{\multirow{2}[4]{*}{\textbf{n}}} & \multicolumn{3}{p{20em}}{\textbf{\%EER}} \\
\cmidrule{2-4}          & \textbf{All Obs.} & \textbf{No Unknown Apps.} & \textbf{No Unforeseen Obs.} \\
    \midrule
    \textbf{20} & \cellcolor[rgb]{ .992,  .918,  .514}43.20 & \cellcolor[rgb]{ .973,  .412,  .42}49.22 & \cellcolor[rgb]{ .976,  .435,  .427}48.96 \\
    \midrule
    \textbf{30} & \cellcolor[rgb]{ .706,  .835,  .498}39.03 & \cellcolor[rgb]{ .996,  .8,  .494}44.72 & \cellcolor[rgb]{ .984,  .631,  .463}46.70 \\
    \midrule
    \textbf{40} & \cellcolor[rgb]{ .569,  .796,  .49}36.97 & \cellcolor[rgb]{ 1,  .894,  .514}43.64 & \cellcolor[rgb]{ .992,  .776,  .49}45.01 \\
    \midrule
    \textbf{50} & \cellcolor[rgb]{ .471,  .769,  .486}35.53 & \cellcolor[rgb]{ .922,  .898,  .51}42.19 & \cellcolor[rgb]{ .996,  .851,  .506}44.16 \\
    \midrule
    \textbf{60} & \cellcolor[rgb]{ .388,  .745,  .482}34.31 & \cellcolor[rgb]{ .941,  .902,  .514}42.47 & \cellcolor[rgb]{ 1,  .922,  .518}43.29 \\
    \bottomrule
    \end{tabular}%
\label{tab:ModEDRes}
\end{table}

In the next section, some suitable modeling approaches for the application usage-based active authentication problem are discussed. 

\section{Suitable Modeling Techniques}\label{sec:VerificationMethods}
In light of the outcomes of the experiments presented in the previous section, it can be asserted that the application-usage-based verification models must be capable of taking into account unknown applications and unforeseen observations. A popular approach to model temporal data sequences is to use state-space models such as Mobility Markov Chains or Hidden Markov Models (HMM) which can model time variation of the data. However, these methods are not capable of handling unforeseen events by default. For example, any unforeseen event will be given a zero emission probability in these models, and therefore, the models will be somewhat like the binary decision model that was discussed earlier. However, simple modifications to these models can improve the usability of these methods when unforeseen events are present as discussed in \cite{PATH_MahbubChellappa2016} for geo-location data. In this paper, the three state-space models namely, the Markov Chain (MC)-based Verification, HMM with Laplacian Smoothing (HMM-lap) and Marginally Smoothed HMM (MSHMM), described in \cite{PATH_MahbubChellappa2016} are employed on the application-usage-based verification task and the performances are compared.  

For the MC method, the prior probability for unknown and unforeseen events are set to a very small nonzero probability of $\delta=e^{-20}$ (Laplace-smoothing) when training a model $X_T$ for observation sequences of length $T$. For MC, the probability of transitioning to an observation state $o_j$ depends only on the probability of the last observation state $o_i$, i.e.
\begin{equation}
\tau_{i,j}=Prob(X_T=o_j|X_{T-1}=o_i).
\end{equation}
If the prior probability of entering any state $i$ is $p_i=Prob\{X_0=i\}$ with respect to the set of observations for user-$z$ $O_T^z$ , then the total probability of traversing any sequence of $n$ consecutive observations $i_0, \hdots, i_n \in O_T^z$ is calculated as
\begin{equation}
Prob(X_0=i_0, \hdots, X_n=i_n)=p_{i_0}\tau_{i_0, i_1}\hdots \tau_{i_{n-1}, i_{n}}
\end{equation}

Similar to the MC method, in HMM-lap method Laplacian Smoothing of the emission probabilities is considered with HMM to incorporate unforeseen observations as discussed in \cite{PATH_MahbubChellappa2016}. The number of hidden states is fixed to $20$ for all the experiments and the maximum number of iteration is set to $50$.

The most suitable approach for handling unforeseen observations is the Marginally Smoothed Hidden Markov Model (MSHMM) introduced in \cite{PATH_MahbubChellappa2016}. To adopt the approach for the active authentication problem, the marginal probabilities of the presence of an application in the training sequence of a user for each time-zone and day flags are precomputed. Assuming that the probability of user-x using application $a_x^i$ at time-zone $TZ(t)$ at time $t$, $P(a_x^i, T_j)$ is independent of the probability of user-x using the application at location $W(t)$, $P(a_x^i, W(t))$ at time $t$, the emission probability from state $s$ to observation $o_t$, $\widehat{e}_{s}(o_t)$ is $P(O_t^{\{a_x, TZ(t), W(t)\}}=o_t^{\{a_x, TZ(t), W(t)\}} |X_t=s)$ if $o_t\in O_t^{\{a_x^p, TZ(t), W(t)\}}$. Otherwise, 
\begin{eqnarray}
\widehat{e}_{s}(o_t) &=& P(O_t^{\{a_x, TZ(t)\}}=o^{\{a_x, TZ(t)\}} |X_t=s)\times \nonumber\\
&& P(O_t^{\{a_x^p, W(t)\}}=o_t^{\{a_x, W(t)\}} |X_t=s), 
\end{eqnarray}
where $P(o_t^{\{a_x, TZ(t)\}}=max(\delta, P(a_x, TZ(t)))$ and, $P(o_t^{\{a_x, W(t)\}}=max(\delta, P(a_x, W(t)))$.
By definition, the MSHMM approach is capable of differentiating between unknown applications and unforeseen observations with known applications, as well as, the more frequent vs. less frequent applications occurring at different time zones and days. 

In the next section, experimental results for these three verification methods are discussed in detail for performance comparison. 

\begin{table*}[t]
  \centering
  \caption{Application-usage-based verification performance comparison for UMDAA-02 dataset across different methods based on EER ($\%$)  for varying sequence length (n) and sampling rate. The number of hidden states is fixed at $20$ and maximum number of iteration is $50$ for HMM-based methods.}
    \begin{tabular}{cp{5.045em}cccccc}
    \toprule
    \multicolumn{1}{c}{\multirow{2}[4]{*}{\textbf{n}}} & \multirow{2}[4]{*}{\textbf{Method}} & \multicolumn{6}{p{24.27em}}{\textbf{Sampling Rate}} \\
\cmidrule{3-8}          & \multicolumn{1}{c}{} & \multicolumn{1}{p{4.045em}}{\textbf{1/5}} & \multicolumn{1}{p{4.045em}}{\textbf{ 1/10}} & \multicolumn{1}{p{4.045em}}{\textbf{1/15}} & \multicolumn{1}{p{4.045em}}{\textbf{1/20}} & \multicolumn{1}{p{4.045em}}{\textbf{1/25}} & \multicolumn{1}{p{4.045em}}{\textbf{1/30}} \\
    \midrule
    \multirow{4}[8]{*}{\textbf{20}} & \textbf{M-ED} & \cellcolor[rgb]{ .98,  .498,  .439}42.96 & \cellcolor[rgb]{ .98,  .498,  .439}42.92 & \cellcolor[rgb]{ .973,  .412,  .42}44.12 & \cellcolor[rgb]{ .976,  .447,  .427}43.64 & \cellcolor[rgb]{ .98,  .486,  .435}43.09 & \cellcolor[rgb]{ .976,  .478,  .435}43.2 \\
\cmidrule{2-8}          & \textbf{MMC} & \cellcolor[rgb]{ .988,  .647,  .467}40.86 & \cellcolor[rgb]{ .988,  .671,  .471}40.53 & \cellcolor[rgb]{ .988,  .686,  .475}40.27 & \cellcolor[rgb]{ .992,  .745,  .486}39.48 & \cellcolor[rgb]{ .988,  .678,  .475}40.39 & \cellcolor[rgb]{ .98,  .914,  .514}36.78 \\
\cmidrule{2-8}          & \textbf{HMM-lap} & \cellcolor[rgb]{ .996,  .816,  .498}38.49 & \cellcolor[rgb]{ .996,  .824,  .502}38.35 & \cellcolor[rgb]{ 1,  .863,  .51}37.82 & \cellcolor[rgb]{ 1,  .894,  .514}37.39 & \cellcolor[rgb]{ .996,  .792,  .494}38.83 & \cellcolor[rgb]{ .98,  .914,  .514}36.77 \\
\cmidrule{2-8}          & \textbf{MSHMM} & \cellcolor[rgb]{ 1,  .898,  .514}37.3 & \cellcolor[rgb]{ 1,  .898,  .514}37.3 & \cellcolor[rgb]{ .973,  .914,  .514}36.67 & \cellcolor[rgb]{ .906,  .894,  .51}35.93 & \cellcolor[rgb]{ .882,  .886,  .51}35.63 & \cellcolor[rgb]{ .808,  .867,  .506}34.82 \\
    \midrule
    \multirow{4}[8]{*}{\textbf{30}} & \textbf{M-ED} & \cellcolor[rgb]{ .98,  .514,  .439}42.7 & \cellcolor[rgb]{ .984,  .584,  .455}41.71 & \cellcolor[rgb]{ .988,  .694,  .475}40.18 & \cellcolor[rgb]{ .996,  .839,  .502}38.17 & \cellcolor[rgb]{ 1,  .878,  .51}37.58 & \cellcolor[rgb]{ .992,  .776,  .49}39.03 \\
\cmidrule{2-8}          & \textbf{MMC} & \cellcolor[rgb]{ .988,  .686,  .475}40.29 & \cellcolor[rgb]{ .992,  .765,  .49}39.18 & \cellcolor[rgb]{ .996,  .835,  .502}38.21 & \cellcolor[rgb]{ .992,  .706,  .478}40 & \cellcolor[rgb]{ .992,  .776,  .49}39.04 & \cellcolor[rgb]{ .984,  .918,  .514}36.82 \\
\cmidrule{2-8}          & \textbf{HMM-lap} & \cellcolor[rgb]{ 1,  .902,  .514}37.28 & \cellcolor[rgb]{ 1,  .906,  .518}37.2 & \cellcolor[rgb]{ .973,  .914,  .514}36.68 & \cellcolor[rgb]{ 1,  .871,  .51}37.73 & \cellcolor[rgb]{ 1,  .859,  .506}37.89 & \cellcolor[rgb]{ 1,  .89,  .514}37.45 \\
\cmidrule{2-8}          & \textbf{MSHMM} & \cellcolor[rgb]{ .933,  .902,  .514}36.23 & \cellcolor[rgb]{ .988,  .918,  .514}36.87 & \cellcolor[rgb]{ .89,  .89,  .51}35.74 & \cellcolor[rgb]{ .988,  .918,  .514}36.87 & \cellcolor[rgb]{ .914,  .894,  .51}35.99 & \cellcolor[rgb]{ .894,  .89,  .51}35.79 \\
    \midrule
    \multirow{4}[8]{*}{\textbf{40}} & \textbf{M-ED} & \cellcolor[rgb]{ .984,  .584,  .455}41.7 & \cellcolor[rgb]{ .996,  .804,  .498}38.64 & \cellcolor[rgb]{ .996,  .82,  .498}38.41 & \cellcolor[rgb]{ .996,  .839,  .502}38.13 & \cellcolor[rgb]{ 1,  .89,  .514}37.45 & \cellcolor[rgb]{ 1,  .922,  .518}36.97 \\
\cmidrule{2-8}          & \textbf{MMC} & \cellcolor[rgb]{ .992,  .757,  .486}39.29 & \cellcolor[rgb]{ .988,  .667,  .471}40.57 & \cellcolor[rgb]{ .996,  .839,  .502}38.13 & \cellcolor[rgb]{ .992,  .733,  .482}39.62 & \cellcolor[rgb]{ .984,  .569,  .451}41.97 & \cellcolor[rgb]{ .902,  .894,  .51}35.89 \\
\cmidrule{2-8}          & \textbf{HMM-lap} & \cellcolor[rgb]{ 1,  .894,  .514}37.37 & \cellcolor[rgb]{ 1,  .859,  .506}37.88 & \cellcolor[rgb]{ .98,  .914,  .514}36.75 & \cellcolor[rgb]{ .918,  .898,  .51}36.07 & \cellcolor[rgb]{ .992,  .773,  .49}39.11 & \cellcolor[rgb]{ .792,  .859,  .502}34.62 \\
\cmidrule{2-8}          & \textbf{MSHMM} & \cellcolor[rgb]{ .859,  .878,  .506}35.4 & \cellcolor[rgb]{ .882,  .886,  .51}35.65 & \cellcolor[rgb]{ .741,  .847,  .502}34.026 & \cellcolor[rgb]{ .773,  .855,  .502}34.4 & \cellcolor[rgb]{ .965,  .91,  .514}36.58 & \cellcolor[rgb]{ .608,  .808,  .494}32.54 \\
    \midrule
    \multirow{4}[8]{*}{\textbf{50}} & \textbf{M-ED} & \cellcolor[rgb]{ .988,  .659,  .467}40.69 & \cellcolor[rgb]{ 1,  .851,  .506}37.98 & \cellcolor[rgb]{ .929,  .898,  .51}36.19 & \cellcolor[rgb]{ .875,  .882,  .51}35.55 & \cellcolor[rgb]{ .875,  .886,  .51}35.58 & \cellcolor[rgb]{ .871,  .882,  .51}35.53 \\
\cmidrule{2-8}          & \textbf{MMC} & \cellcolor[rgb]{ .988,  .682,  .475}40.34 & \cellcolor[rgb]{ 1,  .855,  .506}37.92 & \cellcolor[rgb]{ .996,  .804,  .498}38.67 & \cellcolor[rgb]{ .996,  .918,  .514}36.96 & \cellcolor[rgb]{ .992,  .737,  .482}39.57 & \cellcolor[rgb]{ .698,  .835,  .498}33.56 \\
\cmidrule{2-8}          & \textbf{HMM-lap} & \cellcolor[rgb]{ 1,  .922,  .518}36.97 & \cellcolor[rgb]{ .914,  .894,  .51}36.01 & \cellcolor[rgb]{ .957,  .906,  .514}36.48 & \cellcolor[rgb]{ .8,  .863,  .506}34.72 & \cellcolor[rgb]{ .976,  .914,  .514}36.7 & \cellcolor[rgb]{ .733,  .843,  .502}33.95 \\
\cmidrule{2-8}          & \textbf{MSHMM} & \cellcolor[rgb]{ .91,  .894,  .51}35.95 & \cellcolor[rgb]{ .773,  .855,  .502}34.41 & \cellcolor[rgb]{ .796,  .863,  .506}34.67 & \cellcolor[rgb]{ .596,  .804,  .494}32.41 & \cellcolor[rgb]{ .851,  .878,  .506}35.27 & \cellcolor[rgb]{ .388,  .745,  .482}30 \\
    \midrule
    \multirow{4}[8]{*}{\textbf{60}} & \textbf{M-ED} & \cellcolor[rgb]{ .996,  .8,  .494}38.69 & \cellcolor[rgb]{ .906,  .894,  .51}35.93 & \cellcolor[rgb]{ .855,  .878,  .506}35.32 & \cellcolor[rgb]{ .89,  .886,  .51}35.72 & \cellcolor[rgb]{ .824,  .871,  .506}34.97 & \cellcolor[rgb]{ .765,  .851,  .502}34.31 \\
\cmidrule{2-8}          & \textbf{MMC} & \cellcolor[rgb]{ .996,  .827,  .502}38.33 & \cellcolor[rgb]{ 1,  .886,  .514}37.5 & \cellcolor[rgb]{ 1,  .886,  .514}37.5 & \cellcolor[rgb]{ .996,  .851,  .506}38.01 & \cellcolor[rgb]{ .906,  .894,  .51}35.91 & \cellcolor[rgb]{ .769,  .855,  .502}34.35 \\
\cmidrule{2-8}          & \textbf{HMM-lap} & \cellcolor[rgb]{ .851,  .878,  .506}35.31 & \cellcolor[rgb]{ .867,  .882,  .51}35.48 & \cellcolor[rgb]{ .753,  .851,  .502}34.18 & \cellcolor[rgb]{ .663,  .824,  .498}33.15 & \cellcolor[rgb]{ .918,  .898,  .51}36.05 & \cellcolor[rgb]{ .769,  .855,  .502}34.35 \\
\cmidrule{2-8}          & \textbf{MSHMM} & \cellcolor[rgb]{ .741,  .847,  .502}34.036 & \cellcolor[rgb]{ .82,  .867,  .506}34.92 & \cellcolor[rgb]{ .631,  .812,  .494}32.78 & \cellcolor[rgb]{ .678,  .827,  .498}33.33 & \cellcolor[rgb]{ .765,  .851,  .502}34.3 & \cellcolor[rgb]{ .557,  .792,  .49}31.93 \\
    \bottomrule
    \end{tabular}%
  \label{tab:UMDAAResComp}%
\end{table*}%

\section{Experimental Results} 
\label{sec:ExperimentalResults}
The performances of M-ED, MMC, HMM-lap and MSHMM algorithms for the full test sequences of the UMDAA-02 application usage dataset are shown in Table \ref{tab:UMDAAResComp}, where, the sampling rate has been varied from one sample every $5$ seconds to one sample every $30$ seconds with intervals of $5$ seconds, while the number of previous observations $n$ is varied from $20$ to $60$ with intervals of $10$. It can be seen from the table that with smaller sampling rate and bigger $n$, the EER drops for all the methods. The MSHMM outperforms every other method in every case, which can be attributed to the improved  modeling capability of the method due to marginal smoothing. For a practical verification system, the sampling rate and value of $n$ would determine the latency of decision making. In many cases, a sample every $30$ second might be too late and therefore the system designer should choose these parameters carefully. 

\begin{figure}
\centering
\includegraphics[width = 0.48\textwidth]{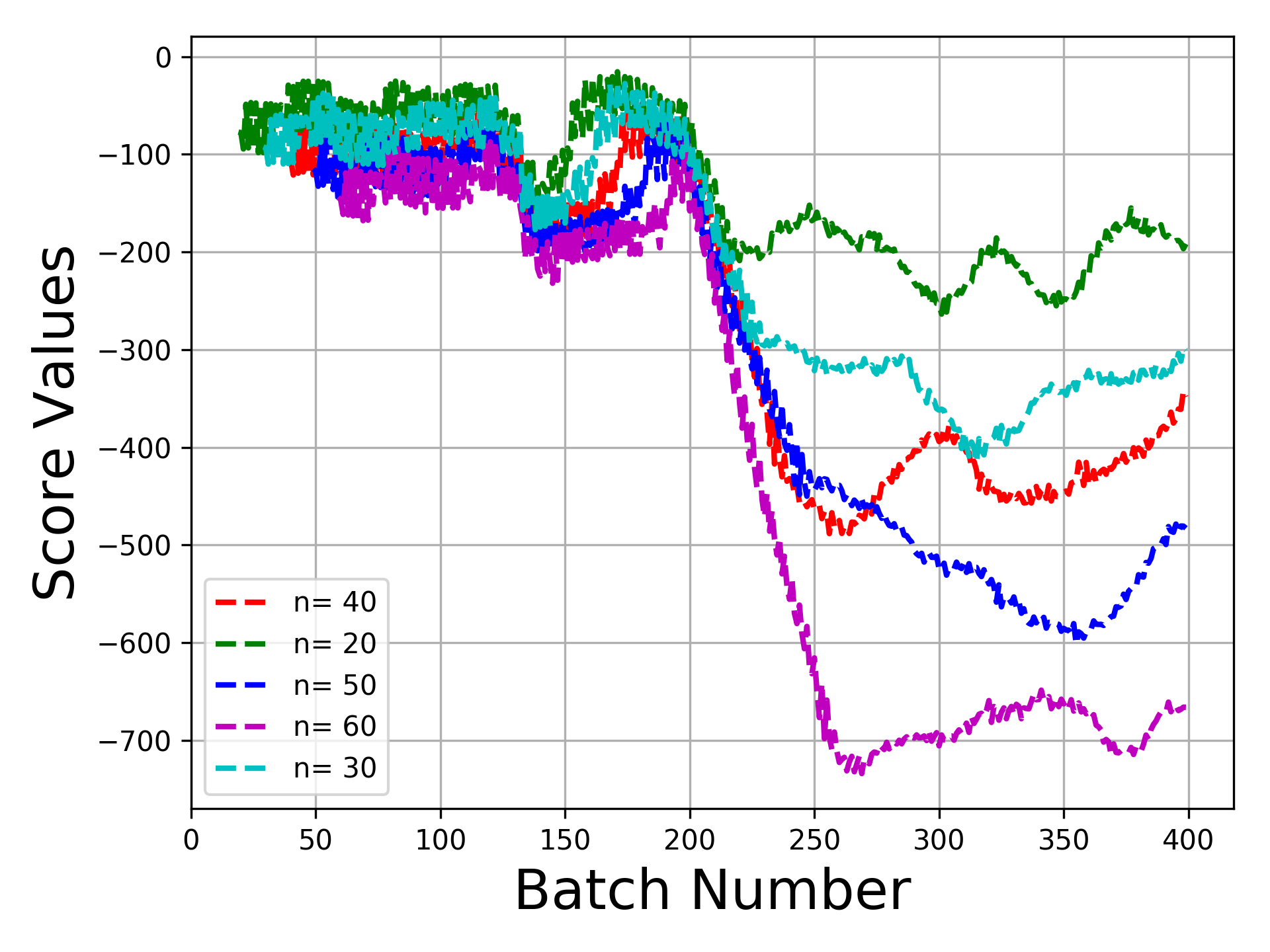}
\caption{Average change in MSHMM scores in response to intrusion on the UMDAA-02 application-usage data. }
\label{fig:AppendAttack_mar_MMC0}
\end{figure}

As for $n$, intuitively with more historical data the performance should improve all the time. In order to determine the impact of $n$ and also to get an idea about the latency of MSHMM when intrusion occurs, a different experiment was performed where a different user's data is appended with the legitimate user's data to simulate intrusion. To be more precise, for each user of the UMDAA-02 dataset, $200$ consecutive observations from the test sequence starting from a random index are appended with $200$ consecutive observations from the test sequences of all the other users (start index picked randomly) and the whole sequence is evaluated using MSHMM for different $n$ values. The average score values across all users are plotted in Fig. \ref{fig:AppendAttack_mar_MMC0} for different $n$ values. When the observations from a different user starts to enter a batch (at $200$-th batch), the average scores returned by MSHMM for each batch drops vividly, as can be seen from the figure. Also, the figure clearly shows the drop is larger for large $n$ values - proving the intuition that considering more historical data is advantageous in this regard. As for latency, if the score of $-200$ is considered as a threshold for decision making, then for all $n=60$, the intrusion will be detected within $\sim 5$ batches, i.e. withing $2.5$ minutes from the inception of intrusion. 

Finally, for the Securacy dataset, the performances of MSHMM, HMM-lap, MMC and M-ED are presented in Table. \ref{tab:SecuracyResults}. Similar to the UMDAA-02 dataset results, MSHMM outperforms the other methods by a good margin. Note that the EER values are much lower for this dataset for the state-space models, which is understandable since it has already been demonstrated in Fig. \ref{AppUsageSimilarityMatrix_UMDAA02_Securacy}(c) that the users are quite separable in this dataset even if only application names are considered. However, M-ED faces difficulty in exploiting the separability of the observations since is not capable of modeling temporal variations as effectively as state-space models.

\begin{table}[htbp]
  \centering
  \caption{App-based verification EER($\%$) comparison for Securacy dataset across different methods \protect\cite{PATH_MahbubChellappa2016} for different $n$ values. Number of Hidden States is set to $20$ and sampling rate is $1/30s^{-1}$.}
    \begin{tabular}{ccccc}
    \toprule
    \textbf{n} & \multicolumn{1}{p{5.045em}}{\textbf{MSHMM}} & \multicolumn{1}{p{5.045em}}{\textbf{MMC}} & \multicolumn{1}{p{5.045em}}{\textbf{HMM-lap}} & \multicolumn{1}{p{5.045em}}{\textbf{M-ED}} \\
    \midrule
    \textbf{20} & \cellcolor[rgb]{ .608,  .808,  .494}17.23 & \cellcolor[rgb]{ 1,  .918,  .518}19.286 & \cellcolor[rgb]{ 1,  .906,  .518}19.66 & \cellcolor[rgb]{ .973,  .412,  .42}35.09 \\
    \midrule
    \textbf{30} & \cellcolor[rgb]{ .51,  .78,  .486}16.75 & \cellcolor[rgb]{ .976,  .914,  .514}18.9967 & \cellcolor[rgb]{ 1,  .906,  .518}19.59 & \cellcolor[rgb]{ .976,  .482,  .435}32.88 \\
    \midrule
    \textbf{40} & \cellcolor[rgb]{ .431,  .757,  .482}16.38 & \cellcolor[rgb]{ .918,  .898,  .51}18.7074 & \cellcolor[rgb]{ 1,  .922,  .518}19.19 & \cellcolor[rgb]{ .98,  .529,  .443}31.4 \\
    \midrule
    \textbf{50} & \cellcolor[rgb]{ .408,  .749,  .482}16.26 & \cellcolor[rgb]{ .761,  .851,  .502}17.9475 & \cellcolor[rgb]{ 1,  .918,  .518}19.22 & \cellcolor[rgb]{ .98,  .561,  .451}30.53 \\
    \midrule
    \textbf{60} & \cellcolor[rgb]{ .388,  .745,  .482}16.16 & \cellcolor[rgb]{ .694,  .831,  .498}17.6443 & \cellcolor[rgb]{ .851,  .878,  .506}18.38 & \cellcolor[rgb]{ .98,  .557,  .451}30.58 \\
    \bottomrule
    \end{tabular}%
  \label{tab:SecuracyResults}%
\end{table}%

Based on results of the experiments presented in this paper, it can be asserted that application-usage data might be useful as a soft biometric for user verification for bolstering the decision in a multi-modal authentication scenario. Given the fact that the application-usage data is readily available and easy to track without using much battery or computational power, real-time score generation is possible. The experiments also depict that the verification scores show rapid change for intrusion within several minutes. Hence, the latency is not too high for a soft biometric measure. However, even though state-space models can be made to work well with some modifications, the equal error rate for a diverse dataset is still around $\sim 16\%$, which needs further improvement. In this regard, bigger training datasets and keeping longer usage history might be helpful. In addition, if computational constraints can be loosened, then more sophisticated high-performance methods such as deep neural networks can be employed to minimize the EER.


\section{Conclusion}\label{sec:conclusion}
In this paper, the challenging problem of active authentication using application usage data has been formulated and systematically tackled to obtain viable solutions. Through several experiments, the impact of unknown applications and unforeseen observations on the authentication problem has been investigated and it is established that for this problem inclusion of the uncertain events are necessary to obtain better performances. In this regard, a modified edit distance algorithm has been introduced, the performance of which is compared with three state-space models namely, Markov Chain, HMM with Laplacian Smoothing and Marginally-Smoothed HMM, in terms of EER. Experiments were performed on the UMDAA-02 and the Securacy application-usage datasets. The experiments revealed some very interesting insights about the differences between the two datasets. Also, the paper addressed different aspects of important practical considerations such as intrusion detection, latency, observation history and sampling rate. As for future work, the M-ED method might be further improved by varying the distances for the three different cases based on the marginal probabilities. Also, recurrent neural network (RNN)-based models might be able to learn more discriminative properties of application-usage patterns. However, RNNs require huge amount of data for useful training, which the two datasets presented here lacks. Another interesting research direction would be the joint training of application sequence and some other sequential data such as the location data to improve the authentication performance. Finally, since application information are also suitable context for other modalities, application data sequences can have duel utilization (as a separate modality and also as context) in more advanced active authentication schemes.

\section*{Acknowledgment}
This work is partially funded by the Emil Aaltonen Foundation (170117 KO), the Finnish Foundation for Technology Promotion (PoDoCo program), Academy of Finland (Grants 286386-CPDSS, 285459-iSCIENCE, 304925-CARE, 313224-STOP, SENSATE), and Marie Skłodowska-Curie Actions (645706-GRAGE)

\bibliographystyle{ieee}
\bibliography{tbiom}
\end{document}